\documentclass[a4paper]{spie}  
\usepackage[]{graphicx}
\usepackage{amssymb,amstext}
\usepackage{cite}
\usepackage{hyperref}

\def\msy{M$_{\odot}$\,yr$^{-1}$}

\def\R23{\mbox{$\rm R_{23}$}}

\title{Planet Formation Imager (PFI): science vision and key requirements
  \footnote{Copyright 2016 Society of Photo-Optical Instrumentation Engineers. One print or electronic copy may be made for personal use only. Systematic reproduction and distribution, duplication of any material in this paper for a fee or for commercial purposes, or modification of the content of the paper are prohibited. DOI: http://dx.doi.org/10.1117/12.2231067}}

\author{Stefan Kraus\supit{a}, 
John D.\ Monnier\supit{b},
Michael J.\ Ireland\supit{c},
Gaspard Duch\^ene\supit{d,e},
Catherine Espaillat\supit{f},
Sebastian H\"onig\supit{g},
Attila Juhasz\supit{h},
Chris Mordasini\supit{i},
Johan Olofsson\supit{j},
Claudia Paladini\supit{k},
Keivan Stassun\supit{l},
Neal Turner\supit{m},
Gautam Vasisht\supit{m},
Tim J.\ Harries\supit{a},\\
Matthew R.\ Bate\supit{a},
Jean-Fran\c{c}ois Gonzalez\supit{n},
Alexis Matter\supit{o},
Zhaohuan Zhu\supit{p},
Olja Panic\supit{h},
Zsolt Regaly\supit{q},
Alessandro Morbidelli\supit{o},
Farzana Meru\supit{h},
Sebastian Wolf\supit{r},
John Ilee\supit{h},
Jean-Philippe Berger\supit{s},
Ming Zhao\supit{t},
Quentin Kral\supit{h},
Andreas Morlok\supit{u},
Amy Bonsor\supit{h},
David Ciardi\supit{v},
Stephen R.\ Kane\supit{w},
Kaitlin Kratter\supit{x},
Greg Laughlin\supit{y},
Joshua Pepper\supit{z},
Sean Raymond\supit{aa},
Lucas Labadie\supit{ab},
Richard P.\ Nelson\supit{ac},
Gerd Weigelt\supit{ad},
Theo ten Brummelaar\supit{ae},
Arnaud Pierens\supit{aa},
Rene Oudmaijer\supit{af},
Wilhelm Kley\supit{ag},
Benjamin Pope\supit{ah},
Eric L.\ N.\ Jensen\supit{ai},
Amelia Bayo\supit{i},
Michael Smith\supit{aj},
Tabetha Boyajian\supit{ak},
Luis Henry Quiroga-Nu\~nez\supit{al,am},
Rafael Millan-Gabet\supit{v},
Andrea Chiavassa\supit{o},
Alexandre Gallenne\supit{an},
Mark Reynolds\supit{b},
Willem-Jan de Wit\supit{an},
Markus Wittkowski\supit{s},
Florentin Millour\supit{o},
Poshak Gandhi\supit{g},
Cristina Ramos Almeida\supit{ao},
Almudena Alonso Herrero\supit{ap},
Chris Packham\supit{aq},
Makoto Kishimoto\supit{ar},
Konrad R.\ W.\ Tristram\supit{an},
J\"org-Uwe Pott\supit{as},
Jean Surdej\supit{at},
David Buscher\supit{h},
Chris Haniff\supit{h},\\
Sylvestre Lacour\supit{au},
Romain Petrov\supit{o},
Steve Ridgway\supit{av},
Peter Tuthill\supit{aw},
Gerard van Belle\supit{ax},
Phil Armitage\supit{ay},
Clement Baruteau\supit{az},
Myriam Benisty\supit{e},
Bertram Bitsch\supit{ba},\\
Sijme-Jan Paardekooper\supit{ac},
Christophe Pinte\supit{e},
Frederic Masset\supit{bb},
Giovanni P.\ Rosotti\supit{h}
\skiplinehalf
\supit{a}University of Exeter, School of Physics, Stocker Road, Exeter, UK;  
\supit{b}Department of Astronomy, University of Michigan, 500 Church St., Ann Arbor, USA;  
\supit{c}Research School of Astronomy \& Astrophysics, Australian National University, Canberra, Australia; 
\supit{d}University of California, Berkeley, USA;
\supit{e}Institut de Plan\'etologie et d'Astrophysique de Grenoble, Grenoble Universit\'e Alpes/CNRS, Grenoble, France;
\supit{f}Boston University, Boston, USA;
\supit{g}University of Southampton, Southampton, UK;
\supit{h}Institute of Astronomy, Madingley Road, Cambridge, UK;
\supit{i}University of Bern, Bern, Switzerland; 
\supit{j}Instituto de Fisica y Astronomia Valparaiso, Chile;
\supit{k}Universite Libre de Bruxelles, Brussels, Belgium;
\supit{l}Vanderbilt University, Nashville, USA;
\supit{m}Jet Propulsion Laboratory, California Institute of Technology, Pasadena/CA, USA;  
\supit{n}Univ Lyon, Univ Lyon1, Ens de Lyon, CNRS, Centre de Recherche Astrophysique de Lyon UMR5574, F-69230, Saint-Genis-Laval, France;
\supit{o}University of Nice-Sophia Antipolis, Observatoire de la C\^ote d'Azur, Nice, France; 
\supit{p}Department of Astrophysical Sciences, Princeton University, Princeton/NJ, USA;   
\supit{q}Konkoly Observatory, Research Center for Astronomy and Earth Sciences, Budapest, Hungary; 
\supit{r}Institute for Theoretical Physics and Astrophysics, University of Kiel, Kiel, Germany;
\supit{s}European Southern Observatory, Karl Schwarschild Strasse, 2, Garching bei M\"unchen, Germany;
\supit{t}Department of Astronomy \& Astrophysics, Pennsylvania State University, PA, USA; 
\supit{u}Institut f\"ur Planetologie M\"unster, M\"unster, Germany;
\supit{v}NASA Exoplanet Science Institute, Pasadena, USA;
\supit{w}San Francisco State University, San Francisco, USA;
\supit{x}University of Arizona, Arizona, USA;
\supit{y}Yale University, New Haven, USA;
\supit{z}Lehigh University, Bethlehem, USA;
\supit{aa}Laboratoire d'Astrophysique de Bordeaux, Bordeaux, France;
\supit{ab}University of Cologne, Cologne, Germany;
\supit{ac}Queen Mary Univeristy of London, London, UK;
\supit{ad}Max Planck Institute for Radioastronomy, Bonn, Germany; 
\supit{ae}Georgia State University, Atlanta, USA;
\supit{af}University of Leeds, Leeds, UK;\\
\supit{ag}University of T\"ubingen, T\"ubingen, Germany;
\supit{ah}University of Oxford, Oxford, UK;\\
\supit{ai}Swarthmore College, Swarthmore, PA USA;
\supit{aj}Kent University, Kent, UK;
\supit{ak}Yale University, New Haven, USA;
\supit{al}Sterrewacht Leiden, Leiden University, Leiden, The Netherlands;
\supit{am}Joint Institute for VLBI ERIC (JIVE), Dwingeloo, The Netherlands;
\supit{an}European Southern Observatory, Santiago, Chile;
\supit{ao}Instituto de Astrofisica de Canarias, La Laguna, Spain; 
\supit{ap}University of Santander/University of Madrid, Madrid, Spain;
\supit{aq}University of Texas, San Antonio, USA;
\supit{ar}Kyoto Sangyo University, Kyoto, Japan;
\supit{as}MPIA, Heidelberg, Germany;
\supit{at}University of Liege, Liege, Belgium;
\supit{au}Observatoire de Paris, Paris, France;
\supit{av}National Optical Astronomy Observatory, Tucson, USA;
\supit{aw}Sydney University, Sydney, Australia;
\supit{ax}Lowell Observatory, Flagstaff, USA;
\supit{ay}University of Colorado and NIST, Colorado, USA; 
\supit{az}Institut de Recherche en Astrophysique et Plan\'etologie, CNRS/Universit\'e de Toulouse, Toulouse, France;
\supit{ba}Lund Observatory, Sweden;
\supit{bb}Universidad Nacional Autonoma de Mexico, Mexico
}

\authorinfo{Further author information: \\
Send correspondence to S.K., E-mail: skraus@astro.ex.ac.uk, Telephone: +44 1392 724125}

\begin{document} 
\maketitle 

\begin{abstract}

The Planet Formation Imager (PFI) project aims to provide a strong scientific vision 
for ground-based optical astronomy beyond the upcoming generation of Extremely Large Telescopes.
We make the case that a breakthrough in angular resolution imaging capabilities 
is required in order to unravel the processes involved in planet formation.
PFI will be optimised to provide a complete census of the protoplanet population 
at all stellocentric radii and over the age range from 0.1 to $\sim 100$\,Myr.
Within this age period, planetary systems undergo dramatic changes
and the final architecture of planetary systems is determined.
Our goal is to study the planetary birth on the natural spatial scale where 
the material is assembled, which is the ``Hill Sphere'' of the forming planet, 
and to characterise the protoplanetary cores by measuring their masses and physical properties.
Our science working group has investigated the observational characteristics of these young protoplanets
as well as the migration mechanisms that might alter the system architecture.
We simulated the imprints that the planets leave in the disk and study how PFI
could revolutionise areas ranging from exoplanet to extragalactic science. 
In this contribution we outline the key science drivers of PFI and discuss the requirements that will guide 
the technology choices, the site selection, and potential science/technology tradeoffs.
\end{abstract}



\section{INTRODUCTION}
\label{sec:intro}

Planet formation is one of the most fascinating and most active areas in 
contemporary astrophysics, linking the field of star formation with exoplanet 
research and with the quest for understanding the origin of our solar system. 

The more than 2000 exoplanetary systems that have been discovered so far
show a surprising diversity in architecture, featuring planet populations such
as the ``Hot Jupiters'' and the ``Super-Earths'' that are not observed in our
solar system.  In order to explain this diversity, theoreticians consider 
different planet formation scenarios and simulate the dynamical mechanisms that
could alter the system architecture following the birth of the planetary cores.

The most commonly discussed planet formation scenarios are the ``core accretion'' 
and the ``gravitational instability'' (GI) scenarios.
Terrestrial planets can only form once the gaseous disk has dissipated, and the 
formation and stability of terrestrial planets depends critically on giant planet formation\cite{mor14}.

In the core accretion scenario \cite{pol96}, the agglomeration of dust grains leads 
to the formation of solid cores of 5-10 Earth masses.
Then the cores start to accrete a significant amount of gas from the protoplanetary disk. 
The gas accretion continues at a roughly constant rate over a few million years, until a 
total mass of 20-30 Earth masses is reached; finally, when the mass in the gas envelope 
is about equal to the core-mass, the gas gravitationally collapses onto the planet: 
the mass of the planet grows exponentially and reaches hundreds of Earth masses in $\sim 10\,000$ years \cite{pol96}.
This model explains well the internal structure of the gas giant planets in our solar system
and is believed to work effectively in the inner few astronomical unit (au) of the disk.  
However, the core accretion model also encounters some fundamental problems -- 
for instance, the planetary cores should undergo a fast orbital migration 
toward the central star ("Type I" migration\cite{war97}) and could be lost before 
having a chance to accrete a massive atmosphere.

\begin{figure}[tb] 
\begin{center}
\includegraphics[width=0.24\textwidth]{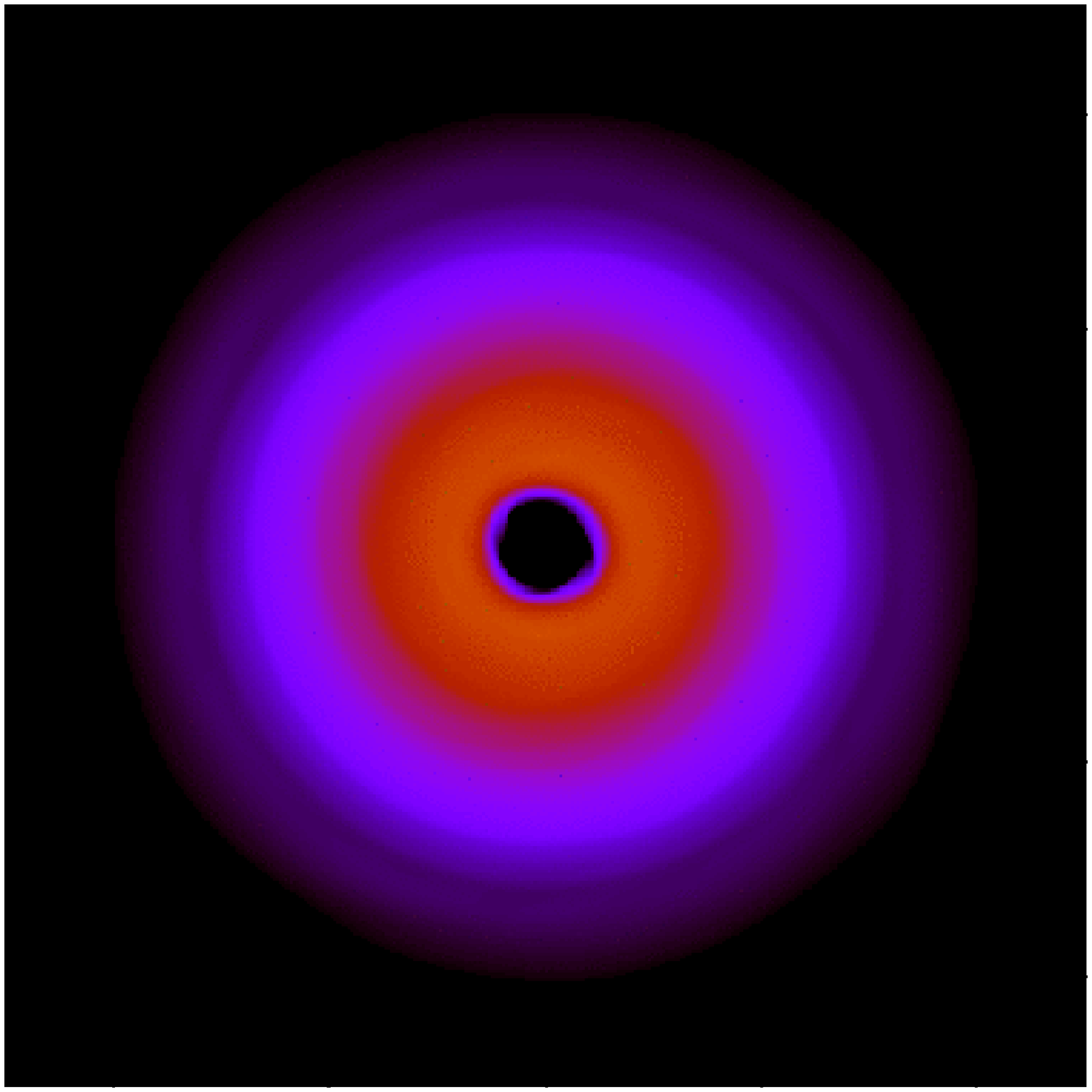}
\includegraphics[width=0.24\textwidth]{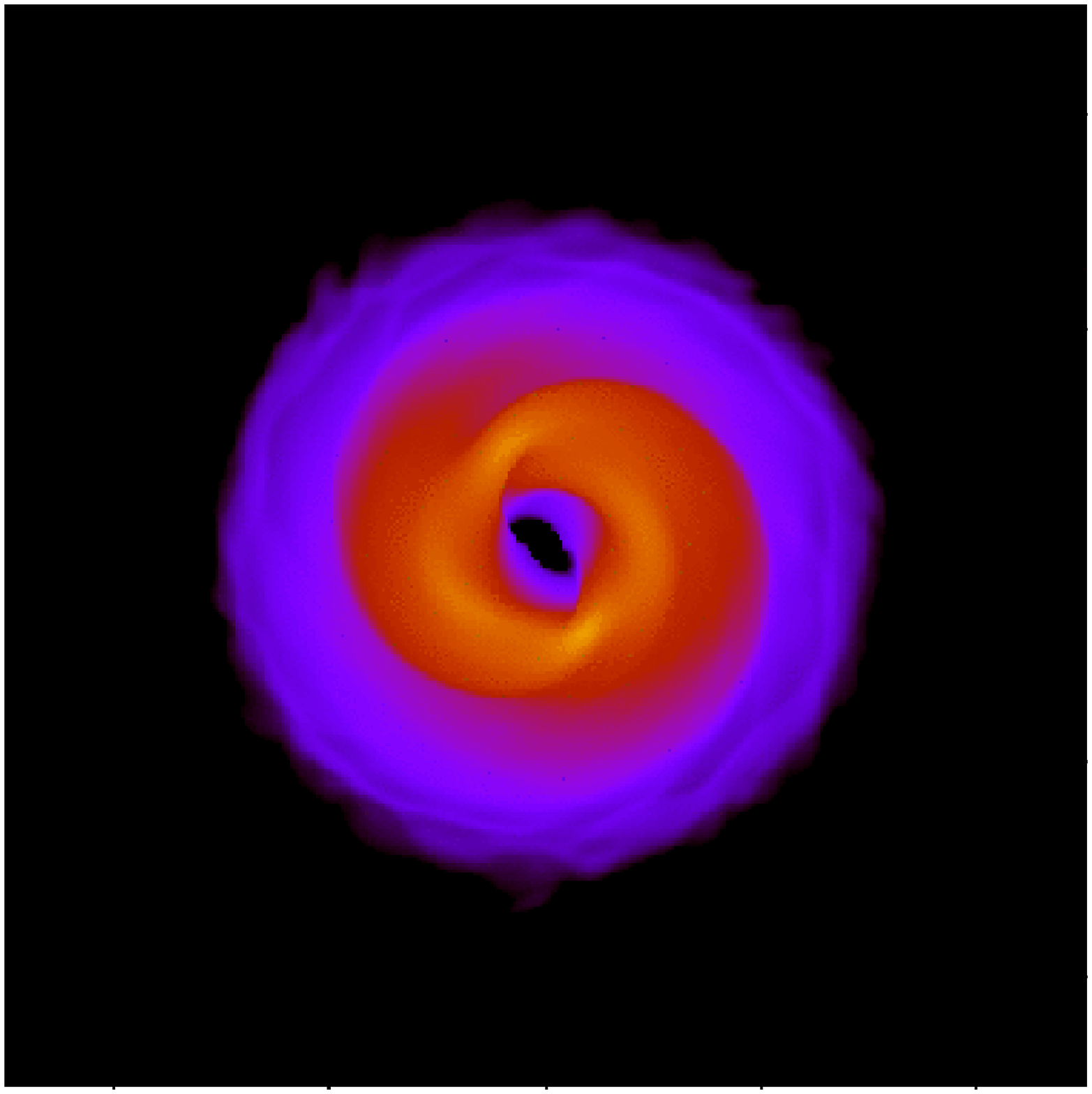}
\includegraphics[width=0.24\textwidth]{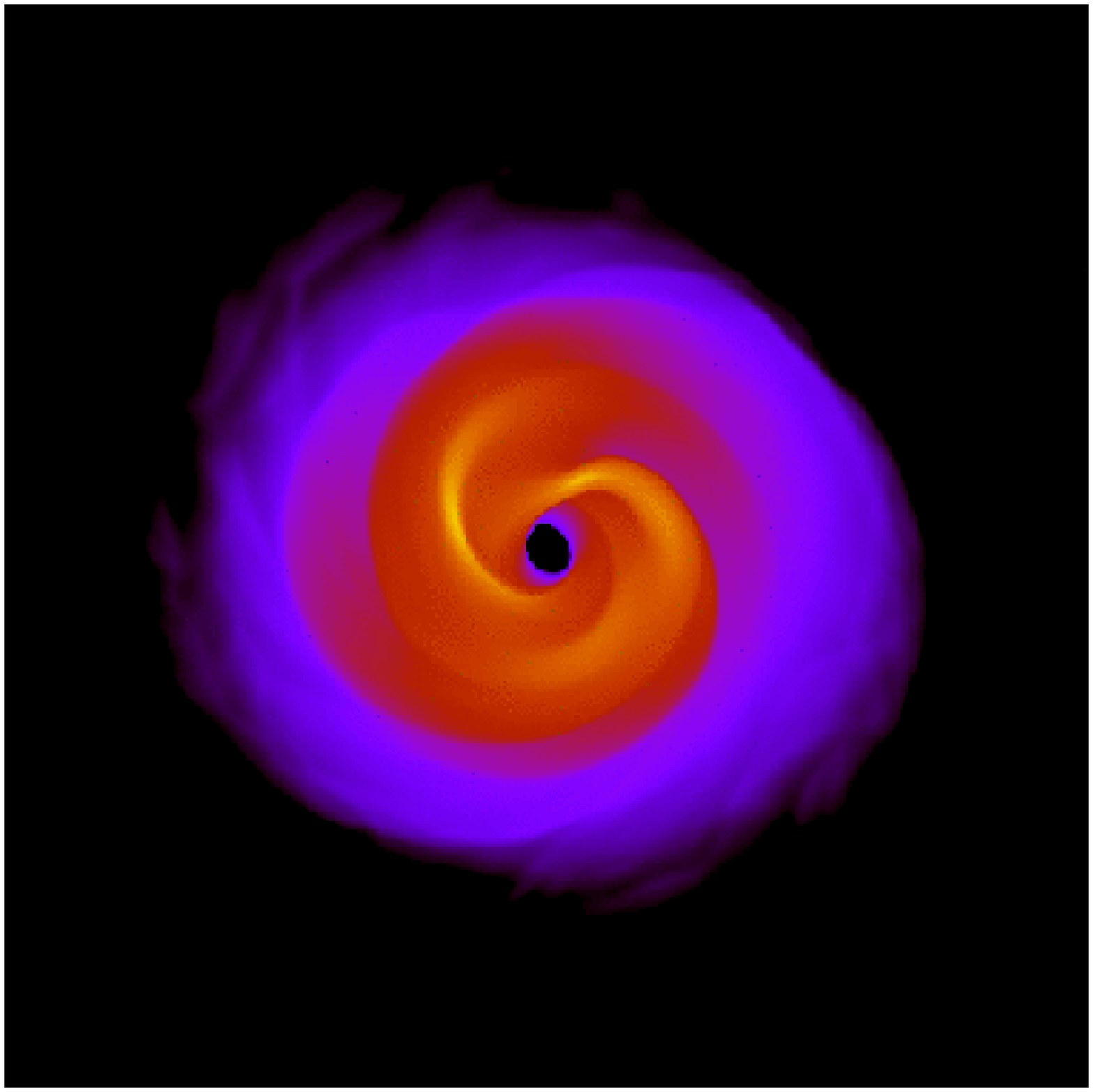}
\includegraphics[width=0.24\textwidth]{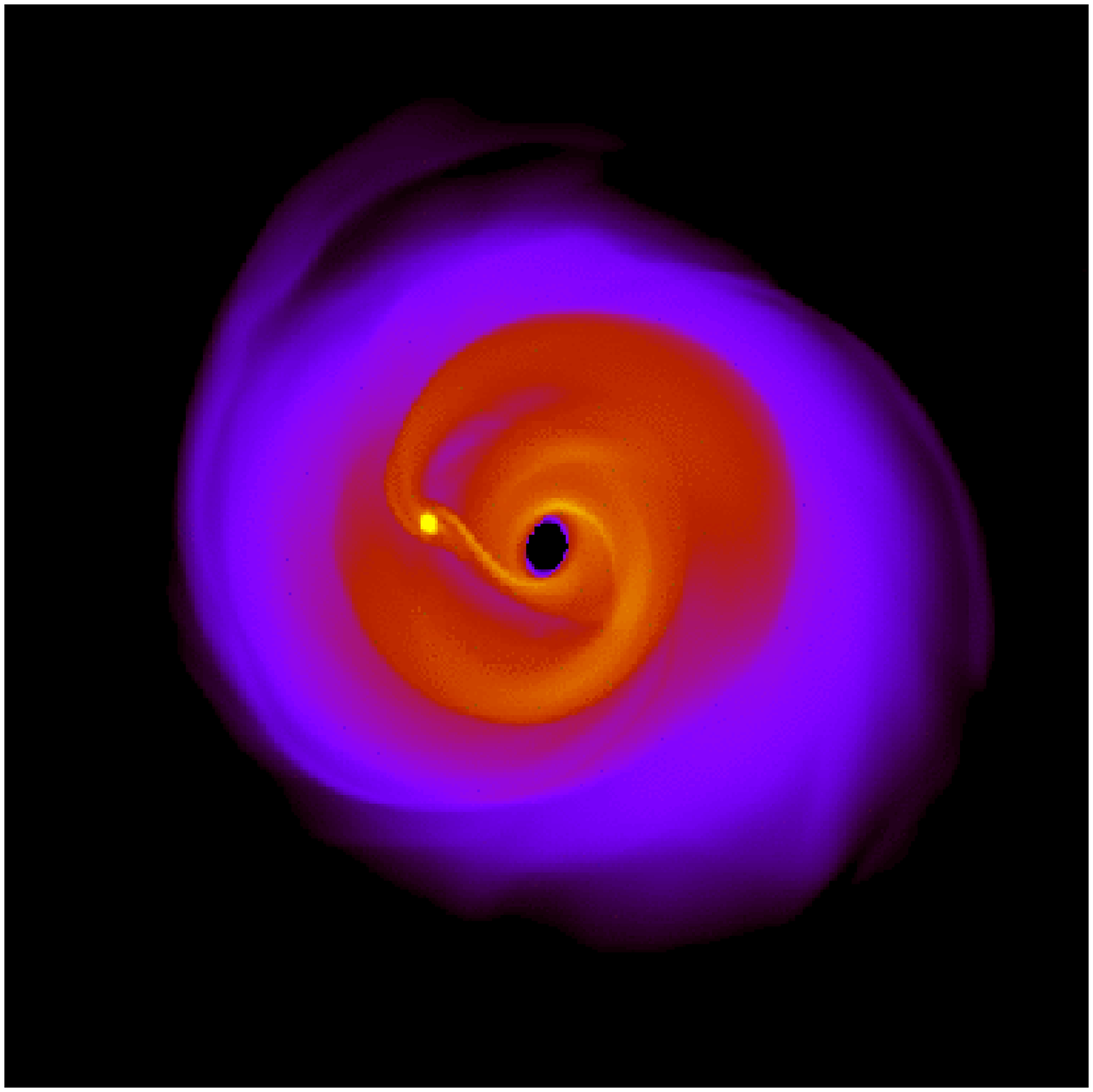}
\caption{Simulation of a gravitationally unstable disk around 
  a  0.3\,M$_{\odot}$ protostar.  
  The simulation covered $\sim 16\,000$ years and were computed
  for a disk with an outer radius of 400\,au.
  The colour  scale indicates  the logarithm  of surface  density, ranging
  from 0.1 -- 300\,g\,cm$^{-2}$.   The disk is gravitationally unstable,
  which leads to formation of spiral waves and causes the disk to 
  collapse into a bound protoplanetary fragment in the final panel.
  Credit: Boley et al.\cite{boley_2009}}
\label{fig:boley_disc} 
\end{center} 
\end{figure}

Planet formation through gravitational instabilities \cite{kuiper_1951, boss_1997,
durisen_2007}, on the other hand, is believed to 
operate primarily in the outer disk regions (on scales of tens to hundreds of AU)
and in the earliest stages of disk evolution, when self-gravity effects can
cause parts of the disk to fragment (Figure~\ref{fig:boley_disc}). 
In contrast to the core accretion model, the key component that is 
important in forming the core is the gas, and hence gaseous
cores form. This may well be the case for Jupiter \cite{saumon_2004},
though  the  capture  of solid  material to form a core after the
fragmenting stage in a gravitationally unstable disk has also been
proposed \cite{helled_2006}.

The goal of the PFI project is to advance our understanding of the 
physical processes involved in planet formation by observing these
processes on the spatial scale where the planets are assembled, which 
is the circumplanetary disk. This poses strong requirements that can
only be achieved with a future telescope facility, whose primary science
goals will be:
\begin{itemize}
\item[(a)] to detect planets at all stello-centric radii and to build a complete picture
of where planets form in disks,
\item[(b)] to characterise the detected protoplanets with spectroscopy and 
to measure their masses from the kinematics in their circumplanetary disk,
\item[(c)] to trace the planet population as function of time in order to understand the influence
of planet-disk interaction mechanisms and other dynamical effects that determine
the final system architecture.
\end{itemize}

Achieving our goals will revolutionize our understanding of the planet formation process,
deepen our understanding of the architecture of exoplanetary systems, and
provide direct insights into the history of our own solar system.
It will allow us to relate exoplanet composition properties to the 
formation and migration history of the planets.

Based on our science objectives it is clear that a key requirement for PFI
is its angular resolution.  We need to resolve the circumplanetary disk 
around the forming planet which scales with the gravitational sphere of influence, or "Hill sphere" $R_{H}$, of the planet. 
For a planet of mass $m$, orbiting a star of mass $M$ on a circular orbit with radius $a$, the radius of the Hill sphere can be approximated as (ref.\cite{ham92})
\begin{equation}
    R_{H} \approx a \left( \frac{m}{3M} \right)^{1/3}.
\end{equation}
Accordingly, the Hill sphere of a Jupiter-mass planet at the location of 
Jupiter in our solar system is 0.35\,au and 0.07\,au for a Jupiter-mass planet
at $a=1$\,au.  Assuming that the target star is located in a nearby star-forming region (distance 140\,pc) and that the circumplanetary disk extends to 
$\sim 0.3 R_{H}$ (ref.\cite{ayl09}), we need to resolve angular sizes of about 0.7\,mas and 0.2\,mas.

These angular scales are one to two orders of magnitude smaller
than the diffraction-limited resolution of the upcoming generation of
Extremely Large Telescopes (ELTs) at near-infrared wavelengths.
Therefore, we plan PFI as an interferometric telescope array that
features kilometric baselines and operates at the relevant infrared wavelengths.
The general science motivation and technology perspective for PFI 
have already been outlined in our 2014 proceeding papers 
(Kraus et al.\cite{kra14b}, Monnier et al.\cite{mon14}, 
Ireland \& Monnier\cite{ire14}).
Therefore we focus in this article on giving a brief update on the 
organisational structure of the Science Working Group (SWG) that we have 
set up and on outlining some initial results and conclusions.  
The activities in the Technical Working Group (TWG) are outlined
in several separate articles in these proceedings (Ireland et al.\cite{ire16}, Monnier et al.\cite{mon16}, Baron et al., 
Mozurkewich et al., Minardi et al., Besser et al., Petrov et al.).

\section{THE PFI SCIENCE WORKING GROUP}
\label{sec:swg}

The PFI science working group is responsible for developing the top-level science requirements of PFI.
In early 2014, we issued an open call to the star \& planet formation, 
exoplanet, and high-angular resolution imaging communities
and invited interested scientists to contribute to our effort.
This call resulted in an overwhelming response of more than 120 scientists that expressed their interest.
More than 70 scientists have already actively contributed towards our science whitebook.

The SWG is structured in 10 groups that investigate different aspects of the 
primary and of auxiliary science drivers.
In the following sections we outlined the topics investigated by our SWG groups,
and present some preliminary results.

The full work will be published as a collection of 10 individual articles 
in a peer-reviewed journal.

\subsection{Protoplanetary Disk Structure \& Disk Physics}

Protostellar disks provide both the raw materials for planet formation 
and the environment where the planets are assembled.  They control when and 
where planets can form, and which elements are available to be incorporated in their cores \& atmospheres.  
Thus one of the main roadblocks to understanding planet formation 
is that we know so little about the basic processes governing the disks' evolution.

This working group investigates fundamental questions related to
disk physics that could be addressed with PFI:
How is orbital angular momentum extracted from accreting material?
What is the nature of the flows in the disks' interiors and atmospheres?
How does the dust evolve over time?
How are outflows launched, accelerated, and collimated?
How are the gas and dust removed as the disks disperse?

The mid-infrared $N$ band (8-13\,$\mu$m) is a particularly interesting
wavelength band for disk studies with PFI, as it provides access
to strong molecular and solid-state spectral features, including the
10~$\mu$m silicate band, several polycyclic aromatic hydrocarbon
bands, and lines of water, OH, carbon monoxide and dioxide, and simple
organic molecules such as HCN and C$_2$H$_2$
(ref. \cite{2008Sci...319.1504C}).  Spatially-resolved imaging in these lines offers
opportunities to probe the radial distribution and transport of the
gas and solid particles \cite{van04}, the chemical reprocessing of
the primordial materials by stellar UV radiation
\cite{2011ApJ...733..102C}, and the mixing between surface layers and
interior \cite{2011ApJ...731..115H}.

We also consider the case for implementing a polarimetric mode
that could map the magnetic field strength in the inner regions of
protoplanetary and/or circumplanetary disks in Zeeman-split spectral lines.

\subsection{Protoplanet Detection}

\begin{figure}[tb]
\centering
\includegraphics[width=0.6\textwidth]{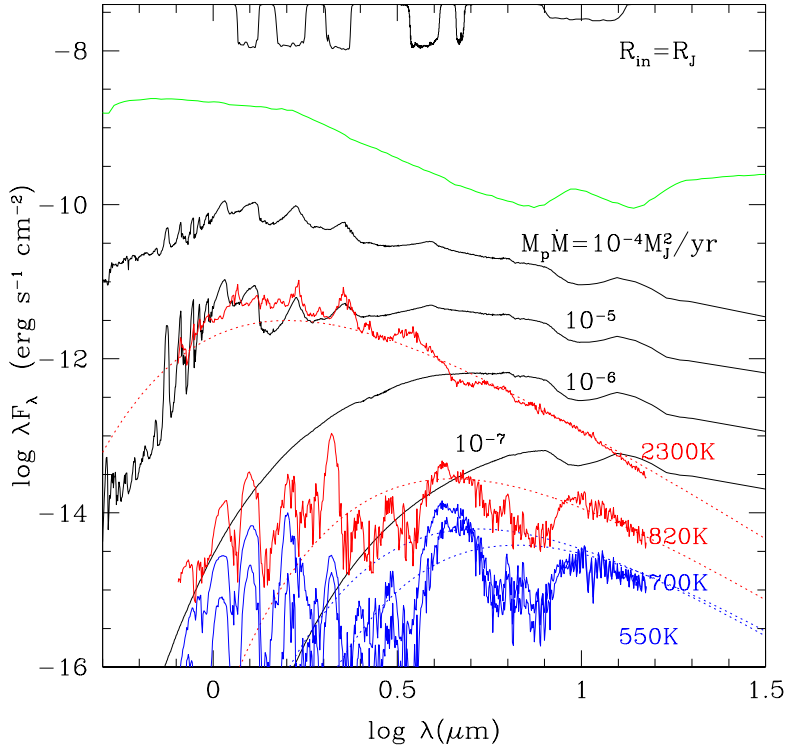} 
\caption{ The SEDs of accreting circumplanetary disks at 100\,pc (black curves).
The product of the planet mass and the disk accretion rate ranges from 10$^{-7}$ to 10$^{-4}$\,M$_{J}^{2}$yr$^{-1}$.
The coloured curves show the SED of a 1\,Myr old planet based on  the ``hot  start'' 
(red curve) and ``cold start'' models (blue curve) \cite{spi12},
where the top curves correspond to a 10 M$_{J}$ planet and the bottom curve for a 1 M$_{J}$ planet.
For another comparison, the green curve is the SED of the protostar 
GM\,Aur (model spectrum from \cite{zhu12}) scaled to 100\,pc. 
At the top of each panel, the black curves indicate the transmission
functions of $J$, $H$, $K$, $L'$, $M$, and $N$ bands. 
Credit: Zhu\cite{zhu15}}
\label{fig:SED}
\end{figure}

This group reviews our knowledge about the observational characteristics of 
young planets, considering both the thermal emission from the planet surface and
of the viscously-heated circumplanetary disk.
Accreting circumplanetary disks can release a large amount of thermal energy due to the 
small size and deep potential of the planet \cite{zhu15}. 
A moderately accreting circumplanetary disk ($\dot{M}\sim 10^{-8}${\msy}; enough to form a 10 M$_{J}$ planet within 1 Myr) 
around a 1 M$_{J}$ planet has a maximum temperature of $\sim$2000 K and an accretion luminosity of
\begin{equation}
L_{disk}=\frac{G{\rm M}_{J}\dot{M}}{2 {\rm R}_{J}}=1.5\times10^{-3} {\rm L}_{\odot}\,,
\end{equation}
which is as bright as a late M-type brown dwarf\cite{bas00,cha00} or a 10 M$_{J}$ planet in a ``hot start'' model
(where ${\rm R}_{J}$ is the Jupiter radius and $\dot{M}$ is the accretion rate on the planet). 
Model spectral energy distributions for circumplanetary accretion disks with different accretion rates are shown in Figure~\ref{fig:SED}.

For spectroscopic observations, CO$_2$, CH$_4$, C$_2$H$_2$, and NH$_3$ all have strong bands 
in the near- and mid-infrared. 
The $M$ band includes the fundamental CO lines that could be suitable for tracing the kinematics of
the circumplanetary disk, although confusion with CO emission from the circumstellar disk might be a problem.
A better line tracer might be H$_2$O, in particular beyond the ice line, where local heating of the accreting
planet might provide the predominant mechanism for producing gas-phase water.
The $L$, $M$, and $N$ bands contain also various line transitions from the Pfund and Humphrey
series that might be suitable for tracing the hot gas in the circumplanetary disk and the
accretion shock onto the planet.

\subsection{Planet Formation Signatures in Pre-Main-Sequence Disks}
\label{sec:pfsignatures}

After a planet has formed it will interact with the parent disk. 
The planet can alter the structure of the surrounding disk for instance by launching
spiral density waves both inside and outside of the planetary orbit. 
Massive planets are capable of opening low density gaps in the disk
or excite local and global disk eccentricity ({Reg{\'a}ly} et al.\cite{reg10,reg14}).
Multiple planets may open several individual gaps or wide common gaps depending on their orbital radius. 
Dust particles may get trapped at pressure maxima at the inner and outer edges of the gap. 
However, the disk influences their embedded planets as well. 
Angular momentum exchange between the planet and the disk results in the change of the 
semi-major axis of the planetary orbit, resulting in the migration of embedded planets. 
The detailed disk properties determine the migration speed
as well as the eccentricity and inclination evolution of the planetary orbits.

Using detailed hydrodynamic simulations, this group investigates 
the observability of planet-disk interactions, for instance through
spiral arms, warps, kinematical signatures, or disk shadows.
We determine how these disk features can be used to infer the
presence and properties of embedded planets, also by combining
PFI imaging with complementary constraints from ALMA or the ELTs.

\subsection{Exoplanetary System Architecture}

\begin{figure}[tb] 
\begin{center}
\includegraphics[width=0.32\textwidth]{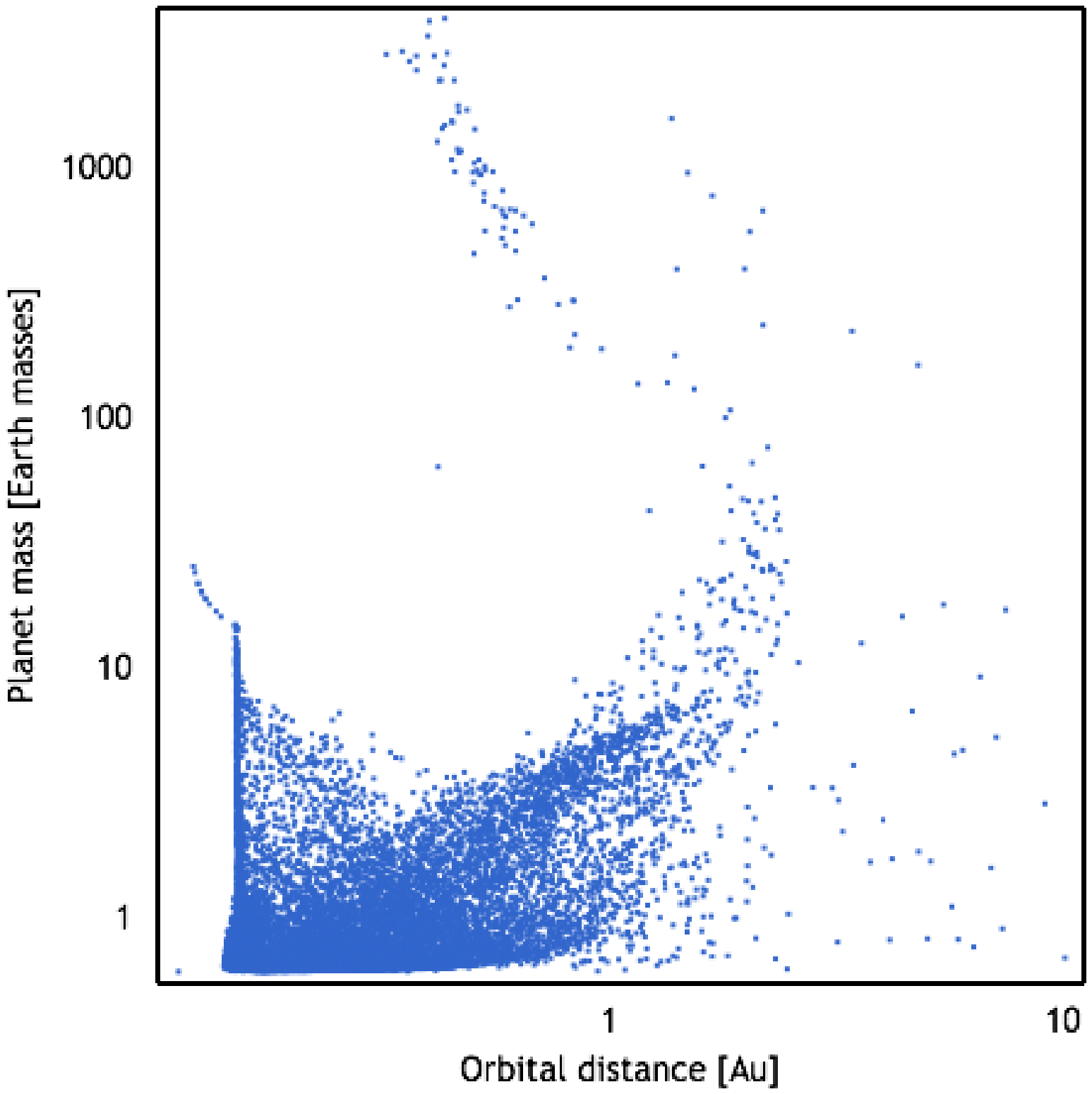}
\includegraphics[width=0.32\textwidth]{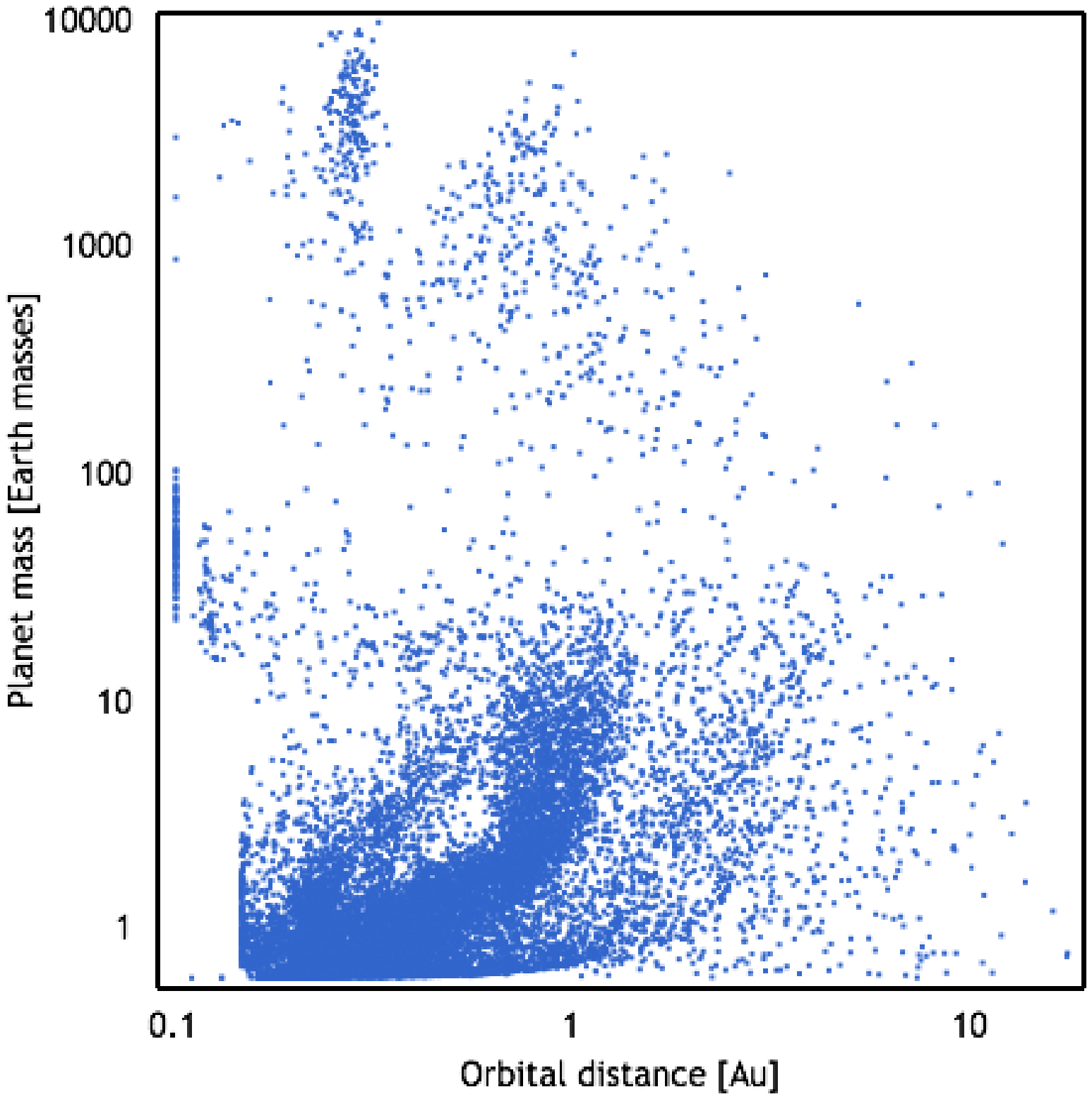}
\includegraphics[width=0.32\textwidth]{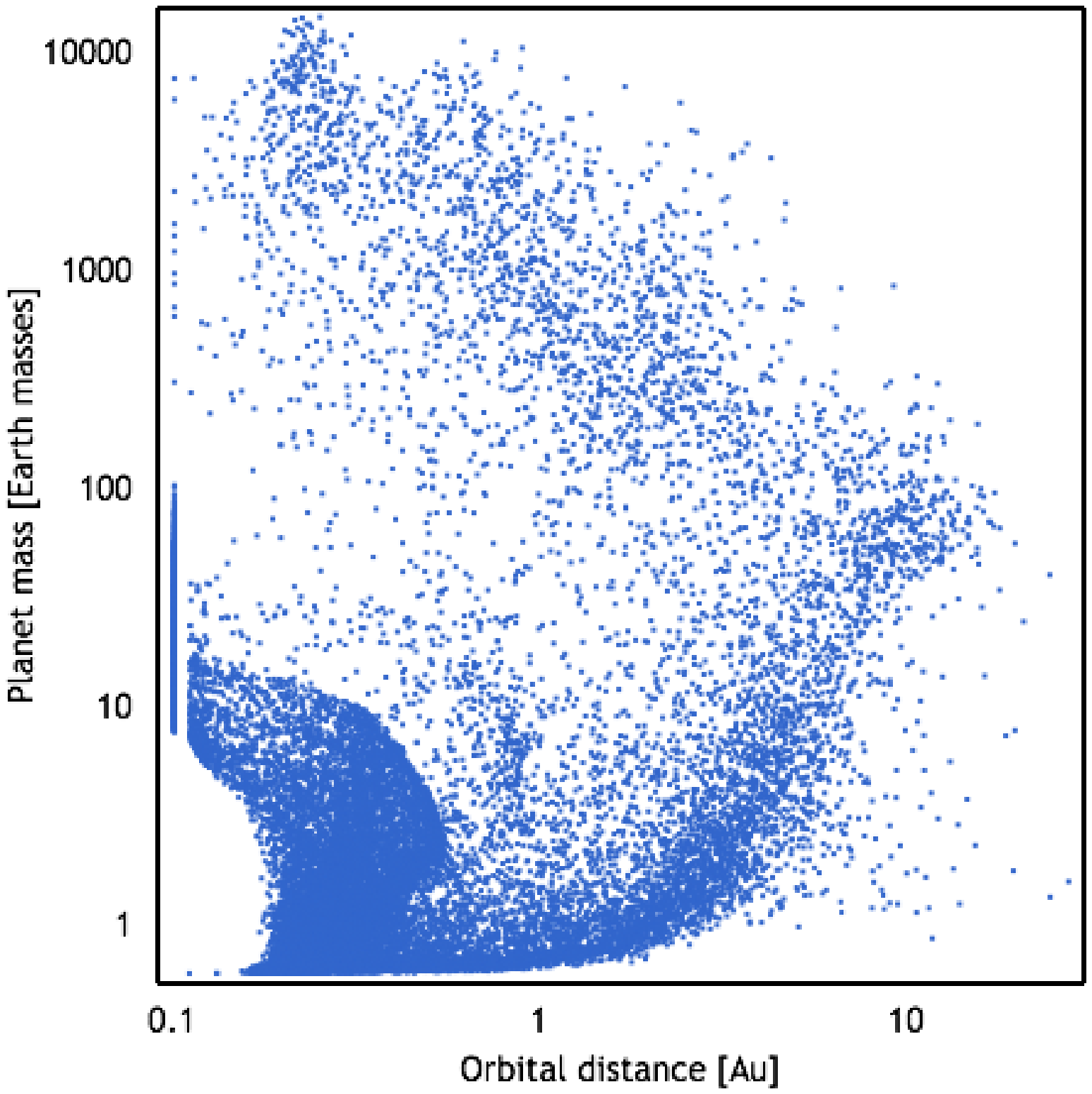}
\caption{
DACE population synthesis models\cite{ali05,mor12,ben14} (population CD753) at time steps of 0.3, 1, and 10\,Myr.  PFI will provide a complete picture of the protoplanet population at these age bins and above a certain mass threshold, which will allow us to much-needed input to constrain these theoretical predictions.
Credit: Data Analysis Center for Exoplanets (DACE), University of Bern (https://dace.unige.ch/evolution/index)}
\label{fig:synthesis} 
\end{center} 
\end{figure}

This group simulates the "initial" exoplanet distribution for different formation 
mechanisms (core accretion, GI, ...) and considers the physical mechanisms 
that are thought to affect the architecture of (exo)planetary systems, 
including inward/outward migration and trapping in 
migration traps (like dead zones or disk truncation points).
We use state-of-the-art population synthesis models to predict how the planet population 
changes during the age range covered by PFI (see Figure~\ref{fig:synthesis}).
Our aim is to identify the main sources of uncertainties in current population 
models and discuss how PFI will be able to capture the changes in 
system architecture as they happen (by observing a statistically significant number 
of systems in different age bins and then comparing the resulting distributions).

\subsection{Planet Formation in Multiple Systems}

Multiplicity is a commonplace property of star formation and planetary systems are found to 
exist in a broad range of multiple systems. 
Binary systems offer a variety of possible configurations, with planets orbiting 
one of the system's components (so called S-type planets) 
or located on a circumbinary orbit (P-type planets). 
Both of these types have been observed by the Kepler spacecraft:
the most compact S-type giant planet-bearing binary system known to date is $\gamma$\,Cep\cite{hat03}, 
whose binary semi-major axis is 20\,au, only 10-times wider than the planet’s orbit. 
Regarding P-type systems, one of the most remarkable Kepler discovery is that of several 
circumbinary planets within mature systems (e.g.\ Kepler 16, ref.\cite{doy11}).
These observations provide a renewed motivation for 
dedicated studies aimed at understanding how the physics of planet formation is affected by the presence of stellar companions. PFI, with its unprecedented imaging fidelity coupled with exquisite interferometric 
resolution will enable studies that directly probe the dynamical processes at play in such an environment. 

PFI will be able to quantify
\begin{itemize}
\item[(a)] the effect that the presence of the stellar companion has on the overall structure of protoplanetary disks,
\item[(b)] how the presence of a stellar companion 
modifies the interaction of a newly-formed planet with its parent disk (planetary accretion, gap opening, migration, eccentricity; e.g.\ Kley et al.\cite{kle08}, {Reg{\'a}ly} et al.\cite{reg12})
\item[(c)] the timescales on which the influence of a stellar companion affects the properties of a newly formed planetary system, and 
\item[(d)] how these effects depend on the mass, semi-major axis, eccentricity, and relative inclination of the binary orbit with respect to the disk.
\end{itemize}
This will provide critical insights on the formation of planetary systems in the context of multiple stellar systems, and how it relates to planet formation around single stars.

\subsection{Late Stages of Planet Formation}

Contemporary studies suggest that the (originally gas-rich) primordial disks will dissipate 
rapidly (e.g. ref.\cite{Hernandez2007,Fedele2010}). The large majority of the gas content 
is believed to be removed from the circumstellar disk by photo-evaporation; 
a pressure driven wind of gas heated by high-energy radiation will efficiently deplete the disk of its gas content, even at large separation (see ref.\cite{Alexander_PPVI} for a recent review). 
The disk depletion leaves a 'debris disk' behind, whose small dust grain population
can be replenished by planetesimal collisions.
What makes debris disks so interesting is not only the origin or their existence but the 
signatures left on the dust from the surrounding environment that can let us probe hidden 
components of the planetary systems. The detection of these signatures at the highest resolution 
and smaller inner working angle would be a priority for PFI. 

Another intriguing science case is to search for giant impact events 
that are believed to mark the late stages of planet formation and
that might also have been involved in the formation of the Earth-Moon system\cite{can01}.
In evolved disks (ages $\gtrsim 100$\,Myr), where most of the circumstellar dust 
has already been dispersed, such dust-creating events should be 
detectable as a strong circumplanetary excess of small dust grains.

\subsection{Star Forming Regions / Target Selection}

The success of PFI will crucially depend on the identification of the 
most promising targets for directly imaging planets during formation. 
In order to build up good statistics on the physics and diversity of planet 
formation pathways, this group seeks to identify a statistically significant number 
of targets in a number of astrophysically relevant ``bins''.
We consider each of these parameters: 
disk structure (i.e.\ tracing the evolution from massive protostellar disks 
to T\,Tauri disks, to transitional disks, and debris disks), stellar age 
(i.e.\ tracing the emergence of planetary system architectures across the 
planet formation epoch from $<$1\,Myr up to $\sim$100\,Myr),
stellar mass (i.e.\ including both low-mass, solar-type T Tauri stars as well as
intermediate-mass Herbig AeBe stars), and star-forming environment 
(i.e.\ including stars in both extreme environments such as dense OB clusters 
and in loose associations).

\begin{figure}[htb] 
\begin{center}
\includegraphics[width=0.45\textwidth]{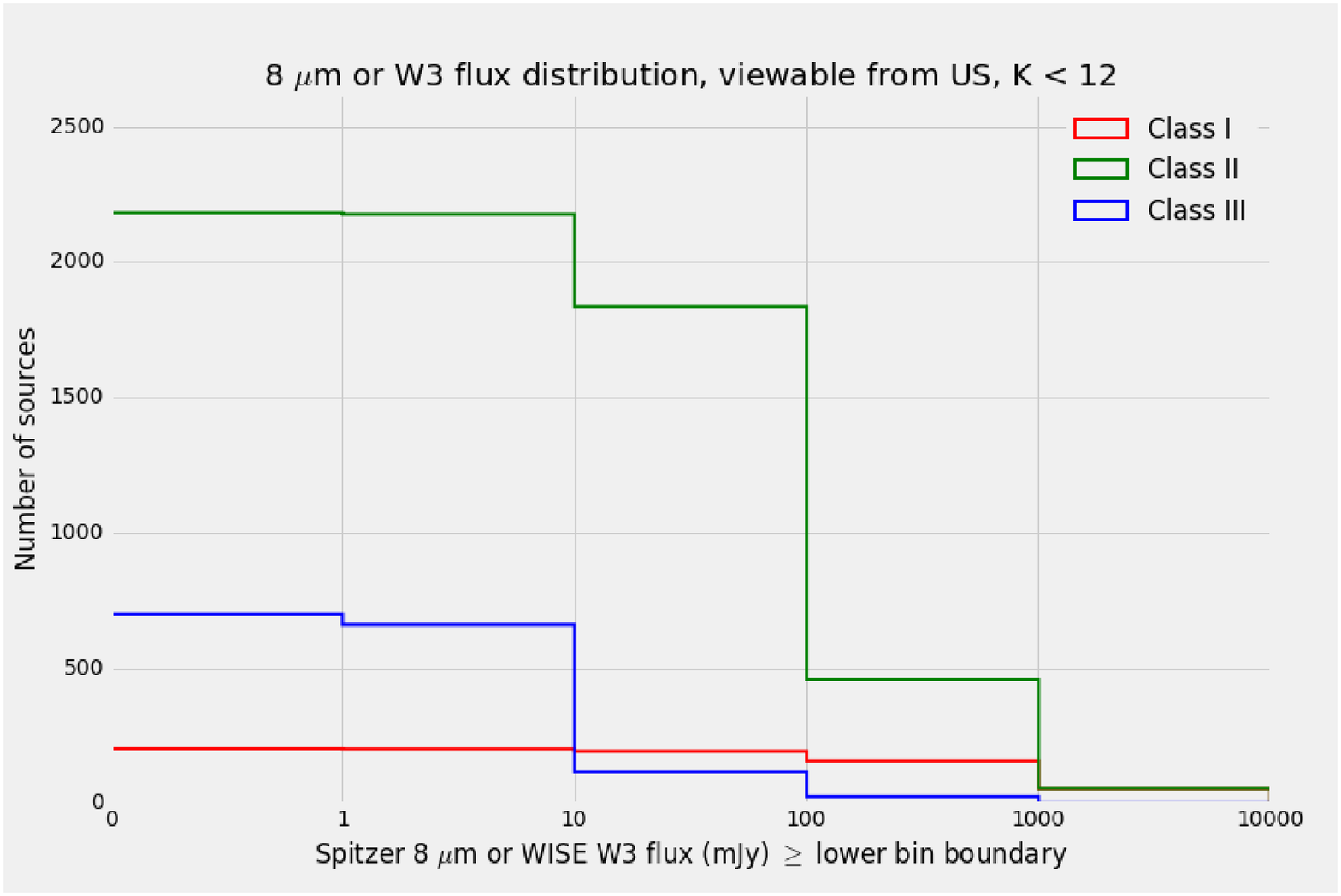}
\includegraphics[width=0.45\textwidth]{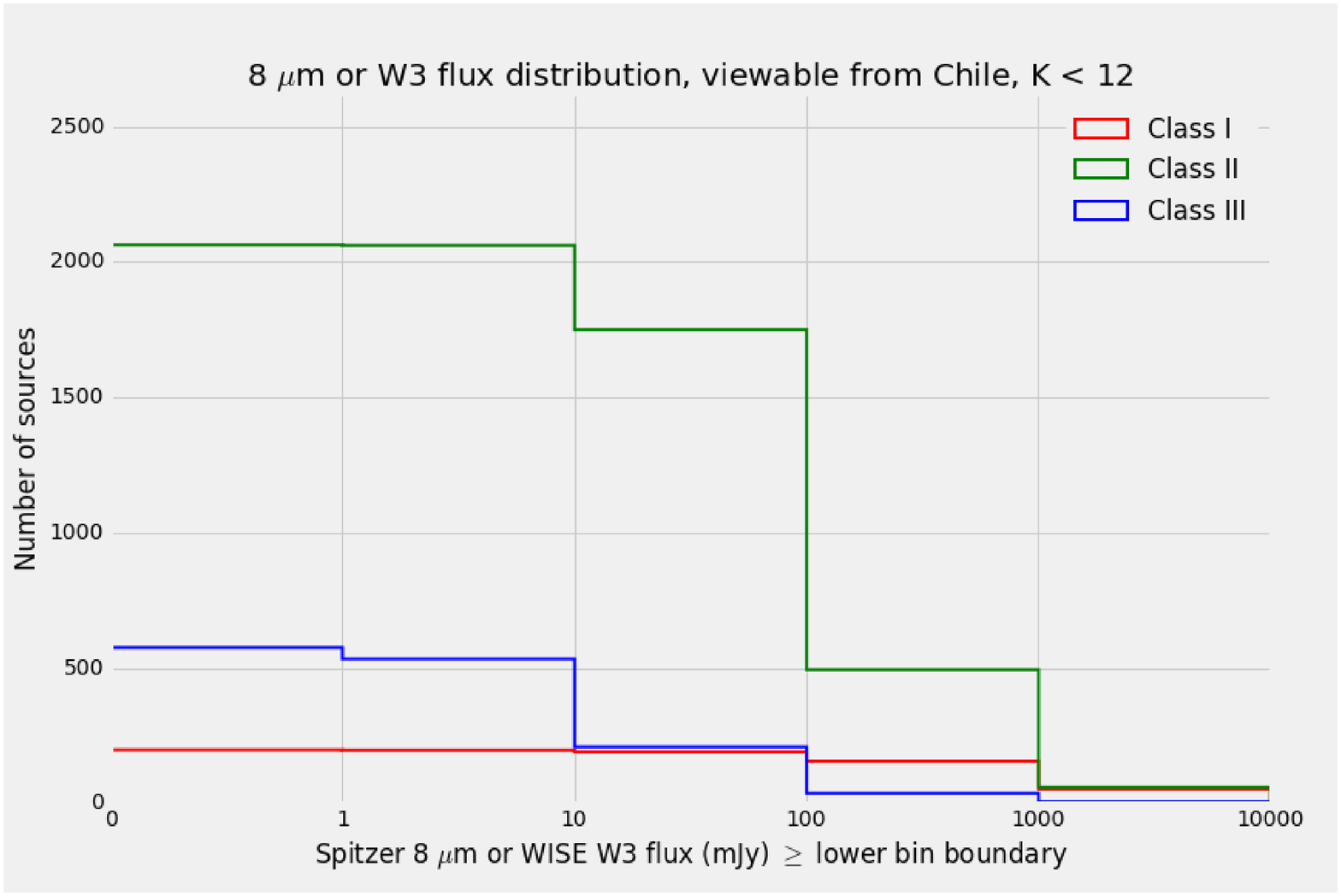}
\includegraphics[width=0.45\textwidth]{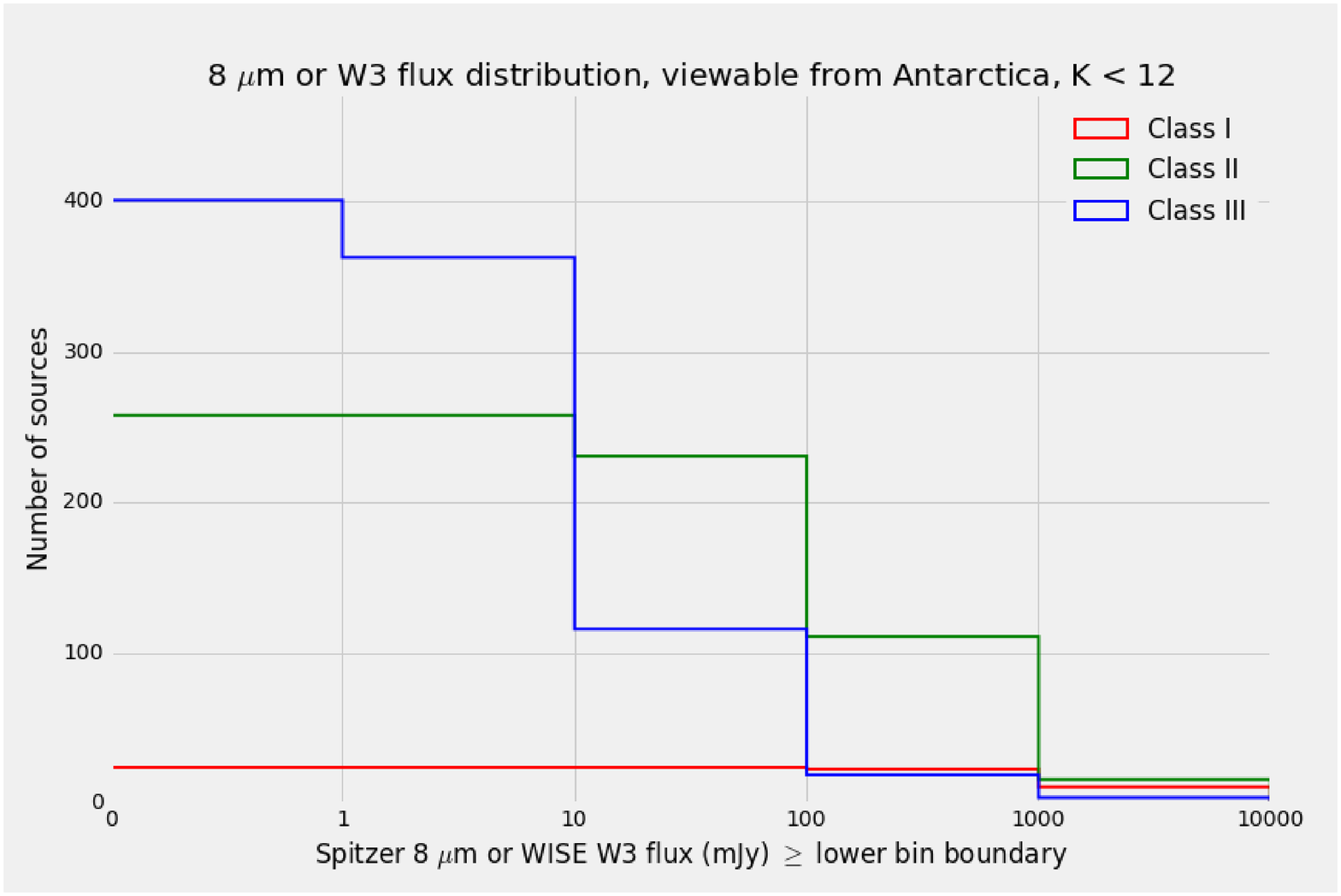}
\caption{Number of targets accessible from a location in the Southern-US (top left), Chile (top right), and
Antarctica (bottom), assuming a fringe tracker limiting magnitude of $K<12$.}
\label{fig:targetsel} 
\end{center} 
\end{figure}

We created a list of 49 star-forming regions that could be targeted by PFI,
including massive clusters, loose associations, and moving groups.  
From this list, we compiled a comprehensive target catalogue 
and determined the number of individual stars that would be accessible as 
function of limiting magnitude for the fringe tracking and science beam combination instrument.
We also considered how the number of potential targets would vary for different
observatory locations, namely in the Southern US, Chile, and Antarctica.
Assuming a sensitivity of K=12\,mag and F$_{8\mu\mathrm{m}}>10$\,mJy (total flux), 
we find that about 2000 targets would be accessible from a Southern-US or Chilean site 
(Figure~\ref{fig:targetsel}, top).
This number is about 5-times lower from a potential Antarctica site 
($\sim 350$; Figure~\ref{fig:targetsel}, bottom), 
due to the worse sky coverage at this extreme latitude.
Therefore, the sensitivity requirements would need to be lowered if one 
would like to benefit from the superior atmospheric conditions at Antarctica.

\subsection{Secondary Science Cases: Exoplanets around Main-Sequence Stars}

PFI will be able to contribute to exoplanet studies by characterising the properties of exoplanet host stars,
both with respect to their fundamental parameters and the characterisation of stellar activity/noise.

Furthermore, we should be able to detect the brightest Hot Jupiters directly and to measure their 
astrometric orbits, which would remove the inclination degeneracy and result in a reliable mass 
determination for bright non-transiting systems, such as 51\,Peg.
The resulting spectra would allow us to determine the atmospheric composition and to 
detect molecular bands, for instance associated with H$_2$O, CO, CO$_2$, and CH$_4$.
We will search for chemical gradients between the day and night side in tidally locked planets
and gain new insights on the presence/absence of clouds and hazes.

\subsection{Secondary Science Cases: Stellar Astrophysics}

With its unique angular resolution, PFI has the potential to deliver breakthroughs in several fields of stellar astrophysics.
Depending on the wavelength band chosen for the final design, the instrument will allow us to image 
the stellar surfaces and/or circumstellar environment of stars at various evolutionary stages. 
The recent studies of the stratification of the dynamic atmosphere of evolved stars (see ref.\cite{wit14} or Paladini in this proceeding) 
are so far limited to nearby objects with solar metallicity. Such kind of works, especially when it comes 
to comparison with model atmospheres, are hampered by distance uncertainties \cite{chi11}. 
A PFI design including baselines of the order of 1-up-to few km will certainly allow us to fully resolve 
giant stars in galactic globular clusters. 
A design including baseline lengths of about 4\,km ($L$ band) would allow to push the observations 
to the Magellanic Clouds. In both cases, by resolving the atmospheres of these giant stars, 
PFI will play a key role in the ongoing discussion of how the mass-loss process acts for stars with different metallicities.

Other key science cases identified involve the study of circumstellar envelopes of Cepheids, as well as the characterisation 
of Cepheids in binary systems; the weather patterns on white dwarf stars, and the accretion of matter onto black-hole in X-ray binaries.

\subsection{Secondary Science Cases: Extragalactic Science \& Cosmology}

There are various fields in extragalactic astrophysics and cosmology that could profit from the high-angular
resolution capabilities provided by PFI. The classical application of infrared interferometry in the field is spatially resolving the parsec-scale dusty environment of active galactic nuclei (AGN). 
We have now a sample of nearly 50 objects for which size and orientation of the dusty region could be constrained. 
However, except for the two brightest objects -- NGC 1068 and the Circinus galaxy -- the current baseline limit 
of 130\,m does not allow us to formally resolve the torus. 
Interestingly, in the few objects with the best intrinsic resolution, we have clear evidence that the distribution of the 
dust in not as expected in the current torus paradigm. 
We rather see a two-component medium, of which the dominant one is suggested to be a dusty wind.
The kilometric baselines of PFI are required to image the brightness distribution and
to separate the different emission components for a large sample of AGN.
For instance, with sensitivity of K=13, approximately 100 AGN at distances out to 1000\,Mpc could be resolved with PFI. 

\begin{figure}[tb] 
\begin{center}
\includegraphics[width=1.0\textwidth]{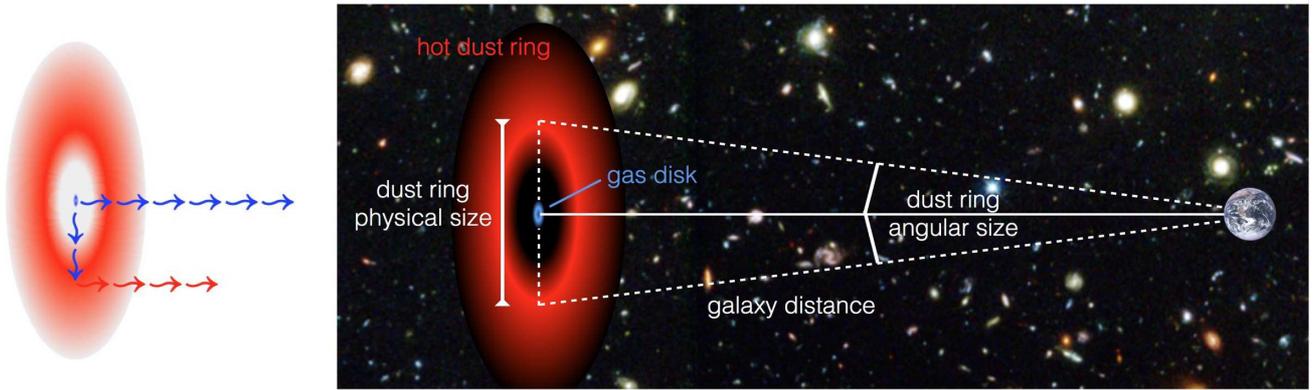}
\caption{Illustration of the "dust parallax distance" measurement method. Using simple geometry, the distance to an AGN can be calculated from the angular size and physical size of the hot dust ring around the supermassive black hole. While reverberation mapping (left) enables us to determine the physical size of the ring (from the time lag between optical and near-infrared emission), the angular size can be measured by infrared interferometers such as PFI (right). Credit: H\"onig et al.}
\label{fig:cosmology}
\end{center} 
\end{figure}

Resolving a large sample of AGN would also open new applications in cosmology,
where the Hubble constant $H_0$ is a key parameter in the
``$\lambda$ Cold Dark Matter ($\lambda$CDM)'' standard model for the state of the universe.
$H_0$ relates the recession velocities of (local) galaxies, or their redshift, 
to their distance. While this relation naturally makes $H_0$ important to determine extragalactic distances from redshifts, the reverse also holds true: 
if we know direct distances to extragalactic sources, we will be able to determine $H_0$. However, these distances estimates are very limited.
The classical approach to measuring distances and, therefore, estimate the Hubble constant uses a set of direct 
and indirect distance indicators, starting from parallax measurements in the Galaxy to Cepheids in satellites and 
nearby galaxies and finally type Ia supernovae (SN\,Ia) further away. This "cosmic distance ladder" hinges on a 
few key direct measurements that anchor the indirect "standard candles" where the luminosity of the source is 
correlated to some observable property (e.g.\ variability period in Cepheids or width of brightness peak of SN\,Ia). 
H\"onig et al.\cite{hon14a} proposed to combine near-infrared continuum reverberation mapping with optical interferometry to measure the geometric distance to AGN.
This method makes use of the fact that the near-IR emission originates within a ring of hot dust at about 
the dust sublimation radius of the dusty torus, which surrounds the black hole and accretion disk in AGN. 
An illustration of the ``dust parallax distance'' method is shown in Figure~\ref{fig:cosmology}. 
H\"onig et al.\cite{hon14b} showed that such distance measurements are very precise ($\sim$10\% uncertainty), 
even if the data are unfavorable, by determining the distance to NGC4151 ($D=19.0^{+2.6}_{-2.4}$\,Mpc). 
PFI could observe a suitably sized sample of AGN that would allow reducing  the  error  to  the  percent  level and make this method a competitor for cosmology.
PFI imaging will remove uncertainties about asymmetric/anisotropic emission in the torus and allow us to image the propagation of the "heating wave" in real time.

\section{CONCLUSIONS}
\label{sec:conclusions}

Since the inception of the project about two years ago, we have set up a 
strong science working group that is investigating the key science drivers for PFI.
The highest science priority for PFI is the detection and characterisation 
of protoplanets during the formation process.
The ELTs will be able to detect protoplanets in the outer disk regions (tens to hundreds of au),
where planets form likely predominantly through gravitational instabilities.
However, the majority of planets is expected to form at smaller separations
(few au) through a fundamentally different mechanism, namely core accretion.
This planet population is inaccessible with ALMA and the ELTs and
PFI will be the only facility that will be able to detect these close-in planets,
revealing the complete protoplanet population at all stellocentric radii.
Most importantly, PFI will be able to trace the evolution of the 
planet population as a function of time and thereby unravel 
the dynamical processes that govern the final architecture of planetary systems.

In order to finalise the technical design for PFI, it is essential to converge
on the optimal observing band, where the $L$/$M$ band or $N$ band are
the most suitable choices.

Our preliminary results indicate that the shorter-wavelength $L$ and $M$ bands are 
optimal for detecting the circumplanetary disk in spectral line tracers.
Line observations offer the prospect to detect even very low-mass
protoplanets and to derive the planet mass from the kinematics of the
circumplanetary disk.  However, given the remaining uncertainties on the
accretion properties of protoplanets, it is currently difficult to predict 
firm detection thresholds.  The line detection method is also more tailored
for detecting young protoplanets, but not able to trace the planets in the
debris disk phase when accretion has ceased.  
These line-detection observations of the protoplanets could likely be achieved 
with a smaller number of apertures and with shorter baseline lengths,
relaxing some of the technical requirements.

The $N$ band, on the other hand, would be the optimal choice for 
tracing the thermal emission of the protoplanets as they cool in the first $\sim 100$\,Myr.
The $10\mu$m dust continuum would also be optimal to image 
planet-induced disk structures and to characterise their dust mineralogy.
However, for continuum detection, the contrast requirements are
higher than for line detections and a large number of apertures
is required to image the complex structures.
The $N$ band hydrogen recombination lines might be suitable for
tracing the circumplanetary accretion disk and the accretion shock onto the planet,
although these lines are weaker than the $L$/$M$ band line tracers.

Based on these considerations we conclude that the science goals
could be achieved either with an $L$/$M$ band or $N$ band-optimised
variant of PFI, where we derive different requirements for these bands 
concerning sensitivity, spectral resolution, contrast, and imaging fidelity.
The TWG will balance our science considerations against the technical readiness
and cost driver considerations when choosing the optimal design.

\acknowledgments     

We acknowledge the work of the many colleagues that contributed to our SWG studies.
Colleagues from all science backgrounds are invited to join our effort and to 
contact us via our project website 
\url{http://www.planetformationimager.org}.\newline
S.K.\ acknowledges support from an STFC Rutherford Fellowship (ST/J004030/1) and
Philip Leverhulme Prize (PLP-2013-110).
Part of this work was carried out at the Jet Propulsion Laboratory, California Institute of Technology, under a contract with the National Aeronautics and Space Administration.
This publication makes use of The Data \& Analysis Center for Exoplanets (DACE), which is a facility based at the 
University of Geneva (CH) dedicated to extrasolar planets data visualisation, exchange and analysis. 
DACE is a platform of the Swiss National Centre of Competence in Research (NCCR) PlanetS, federating the Swiss expertise in Exoplanet research. 
The DACE platform is available at https://dace.unige.ch.


\bibliography{PFIscience}   
\bibliographystyle{spiebib}   

\end{document}